# The Evolution of Multicomponent Systems at High Pressures: IV. The Genesis of Optical Activity in High-density, Abiotic Fluids.

**J. F. Kenney**
*Joint Institute of the Physics of the Earth, Russian Academy of Sciences;*
*Gas Resources Corporation,*
11811 North Freeway, fl. 5, Houston, TX 77060, U.S.A.; JFK@alum.MIT.edu ?
**Ulrich K. Deiters**
*Department of Physical Chemistry, University of Cologne*
Luxemburger Strasse 116, D-50939, Cologne, GERMANY

Abstract:

A thermodynamic argument has been developed which relates the chirality of the constituents of a mixture of enantiomers to the system excess volume, and thereby to its Gibbs free enthalpy. A specific connection is shown between the excess volume and the statistical mechanical partition function. The Kihara-Steiner equations, which describe the geometry of convex hard bodies, have been extended to include also chiral hard bodies. These results have been incorporated into an extension of the Pavlícek-Nezbeda-Boublík equation of state for convex, aspherical, hard-body systems. The Gibbs free enthalpy has been calculated, both for single-component and racemic mixtures, for a wide variety of hard-body systems of diverse volumes and degrees of asphericity, prolateness, and chirality. The results show that a system of chiral enantiomers can evolve to an unbalanced, scalemic mixture, which must manifest optical activity, in many circumstances of density, particle volume, asphericity, and degree of chirality. The real chiral molecules fluorochloroiodomethane, CHFClI, and 4-vinylcyclohexene, $C_8H_{12}$, have been investigated by Monte Carlo simulation, and observed to manifest, both, positive excess volumes (in their racemic mixtures) which increase with pressure, and thereby the racemic-scalemic transition to unbalanced distributions of enantiomers. The racemic-scalemic transition, responsible for the evolution of an optically active fluid, is shown to be one particular case of the general, complex phase behavior characteristic of "closely-similar" molecules (either chiral or achiral) at high pressures.

[Keywords: optical activity, chirality, hard-body systems, thermodynamics.]

## 1.  Introduction.

The phenomenon of optical activity in fluids, either biotic or abiotic, requires, simultaneously, two distinct phenomena: the presence of chiral molecules, which lack a center of inversion; and an *unequal* distribution of those chiral enantiomers. A system of chiral molecules characterized by a distribution of equal abundances of enantiomers is a racemic mixture; ones characterized by distributions of unequal abundances are scalemic mixtures. Only scalemic mixtures manifest optical activity. Certain biological processes, such as natural fermentation, generate chiral molecules of only one enantiomer. Abiological processes can produce either equal or unequal enantiomer abundances, depending upon the thermodynamic conditions of their evolution.

The phenomenon of optical activity in abiotic fluids is shown in the following sections to be a direct consequence of the chiral geometry of the system particles acting according to the laws of classical thermodynamics. In the following sections a purely thermodynamic argument is developed which relates the evolution of optical activity in a system of chiral molecules to the excess volume of scalemic mixtures. The excess volume of the scalemic mixture of enantiomers is related to their geometric properties using the Kihara-Steiner equations, which have been extended to describe particles which lack a center of inversion. The chiral property described by the extension of the Kihara-Steiner equations is introduced into the Pavlícek-Nezbeda-Boublík equations for mixtures of hard bodies, with which are calculated the Gibbs free enthalpies and thermodynamic Affinities of hard-body systems. The calculated thermodynamic Affinities establish that, in accordance with the dictates of the second law, a system of chiral molecules will often evolve unbalanced (scalemic) abundances of enantiomers at high densities.

For experimental verification of the theoretical calculations made with convex hard-body systems, Monte Carlo simulations have been carried out on the real chiral molecules fluorochloroiodomethane, CHFClI, and 4-vinlycyclohexene, $C_8H_{12}$. Both CHFClI, and $C_8H_{12}$ have been observed, at high pressures, to manifest higher densities in their scalemic distributions, as compared to their racemic ones. Such density change drives the racemic-scalemic transition.

### 1.1  Historical background.

From its first demonstration by Pasteur, the phenomenon of optical activity in fluids has engaged the attention and interest of the scientific community.[1] This phenomenon has provided an arena for considerable exercise of imagination and crea-tive fantasy, regrettably almost entirely unleavened by the discipline of thermodynamics.

Perhaps for reason of its historical provenance in fermented wine, the phenomenon of optical activity in fluids was for some time believed to have some intrinsic connection with biological processes or materials. Such error persisted until the phenomenon of optical activity was observed in material, some believed previously to be uniquely of biotic origin, extracted from the interiors of meteorites.

From the interiors of carbonaceous meteorites have been extracted the common amino-acid molecules alanine, aspartic acid, glutamic acid, glycine, leusine, proline, serine, threonine, as well as the unusual ones $\alpha$-aminoisobutyric acid, isovaline, pseudoleucine.[2-4] At one time, all had been considered to have be solely of biotic origin. The ages of the carbonaceous meteorites were determined to be 3-4.5 billion years, and their origins clearly abiotic. Therefore, those amino acids had to be recognized as compounds of both biological and abiological genesis. Furthermore, solutions of amino acid molecules from carbonaceous meteorites were observed to manifest optical activity. Thus was thoroughly discredited the notion that the phenomenon of optical activity in fluids (particularly those of carbon compounds) might have any intrinsic connection with biotic matter. Significantly, the optical activity observed in the amino acids extracted from carbonaceous meteorites has not the characteristics of such of common biotic origin, with only one enantiomer present; instead, it manifests the characteristics observed in natural petroleum, with unbalanced, so-called scalemic, abundances of chiral molecules.

The optical activity commonly observed in natural petroleum was for years speciously claimed as a "proof" of some connection with biological detritus, - albeit one requiring both a willing disregard of the considerable differences between the optical activity observed in natural petroleum and that in materials of truly biotic origin, such as wine, as well as desuetude of the dictates of the laws of thermodynamics. Observation of optical activity, typical of such in natural petroleum, in hydrocarbon material extracted from the interiors of carbonaceous meteorites, discredited those claims.[5, 6] Nonetheless, the scientific conundrum remained as to why the hydrocarbons manifest optical activity, in both carbonaceous meteorites and terrestrial crude oil.

There is common misunderstanding that the molecular property of chirality, which is responsible for optical activity in fluids, is an unusual, complicated property of large, complex, multi-atomic molecules. The small, common, single-branched alkane molecules are usually chiral. Single-branched alkanes comprise between 7-15% of the molecular components of natural terrestrial petroleum and are also observed in petroleum synthesized by such as the Fischer-Tropsch processes. When these chiral molecules are created in low-pressure industrial processes, they occur always in equal enantiomer abundances, and the resulting synthetic petroleum does not manifest optical activity. In natural petroleum, these molecules occur in unequal enantiomer abundances, and the fluid manifests optical activity. Molecular chirality results from the highly directional property of the covalent bond, and is indifferent to whether a compound results from a biological or an abiological process.

Previous hypotheses offered to explain optical activity of the compounds extracted from carbonaceous meteorites have invoked such *deus ex machina* as "panspermia," – the "seeding" of optically active biotic molecules from (literally) the heavens,[7, 8] - or cumulative effects of the chiral weak-interaction involved in beta decay,[9-16] - with necessary oversight of the several orders of magnitude of energy difference compared to that attributable to the entropy of mixing, which would destroy any imbalance responsible for optical activity.

With no recourse to any such artifices, the phenomenon of optical activity in fluids is here shown to be an inevitable consequence of the dictates of thermodynamic stability theory manifested by systems which contain quite ordinary, covalent-bonded molecules, in certain conditions of density.

### 1.2 The organization of this paper.

The topic of optical activity in multicomponent fluids is taken up thoroughly, in order that its fundamental thermodynamic property be set forth explicitly, and that its statistical mechanical genesis be demonstrated. This paper is organized into three parts:

1. A thermodynamic argument is developed which relates the Gibbs free enthalpy, and the phase stability of a mixed system, to its excess volume. This argument invokes no specific molecular property, and uses only a strict thermodynamic definition of a system containing chiral components which specifies equality of chemical potentials and molar volumes, and a non-vanishing excess volume. The distribution of species in a multicomponent system is shown to be determined, at low pressures, by its entropy of mixing, and, at high pressures, by its excess volume.

2. A statistical mechanical argument is developed which relates directly the Gibbs free enthalpy and excess volume to specific molecular geometric properties. The Kihara-Steiner equations have been extended to describe hard-body particles which do not possess a center of inversion; and the results have been applied to the Pavlícek-Nezbeda-Boublík equations for convex hard-body fluids.

3. Using the Pavlícek-Nezbeda-Boublík equations, the thermodynamic Affinity has been calculated formally for a diverse group of hard-body fluid systems characterized by different molecular volumes, and degrees of asphericity and chirality.  These fluids are shown to undergo the racemic-scalemic transition exactly as required by the general thermodynamic argument developed in section 2.

Using Monte Carlo simulation, the two real, chiral fluids, fluorochloroiodomethane, CHFClI, and 4-vinlycyclohexene, $C_8H_{12}$, have been investigated as pure chiral components and as racemic mixtures.  The latter are shown to develop positive excess volumes at increased densities, which increase approximately linearly with pressure, and which therefore induce the racemic-scalemic transition.

## 2      Molecular chirality.

There is common misunderstanding that the molecular property of chirality, which is responsible for optical activity in fluids, is an unusual, complicated property of large, complex molecules, themselves probably of biotic origin and comprised of numerous different atomic species.  This misunderstanding appears supported by such (otherwise erudite) treatises on the subject as that by Jacques *et al.*[17], which discuss such complex molecules as D-glucopyranose, and L-hydroxy-2-hydridamine-d-α-bromocamphor-π-sulfonate, but fail even to mention the chirality of the banal, simple molecules described below.  The purpose of this short section is to correct such misperception with a simple, clear, elementary description of molecular chirality, demonstrated by small, common molecules, comprised of only two atomic species.

Chiral geometry, characterized by many compounds, results from the highly directional property of the covalent bond, itself the characteristic responsible for the wealth and diversity of stereochemistry.  The directional property of the covalent bond, including particularly that of carbon, is indifferent to whether a compound results from either a biological or abiological process.

Consider, for example, the tetrahedral methane molecule, altered so that the hydrogen atom at its uppermost apex remains while those at the 2-, 6-, and 10-o'clock positions on the base of the tetrahedron are replaced, respectively, by a propyl radical (-$C_3H_7$), a methyl radical (-$CH_3$), and an ethyl radical (-$C_2H_5$).  The molecule which results, represented schematically in Fig. 1, is 3-methylhexane, a common, single-branched alkane, observed both in natural petroleum and also in petroleum industrially synthesized by Fischer-Tropsch processes.  Plainly, 3-methylhexane is a chiral molecule.  Equally plainly, the isomer which would result by exchanging the ethyl and methyl radicals, is another, distinct chiral molecule.  [These isomers are often designated (*R*)-3-methylhexane and (*S*)-3-methylhexane, respectively.]  Furthermore, if the propyl radical at the 2-o'clock position of the base of the tetrahedron were substituted by a butyl group (-$C_4H_9$), or any n-alkyl group for which $n > 3$ (-$C_nH_{2n+1}$), the resulting single-branched alkane would also be chiral. Furthermore still, the distinct, single-branched, alkane isomers formed by changing the position of the methyl group from the 3-position to the 2-, or 4-, or any other position on the alkane chain (except the center one), will also be chiral.

These simple considerations demonstrate that chirality is an inevitable and banally common molecular property, particularly among carbon compounds. Single-branched, *chiral* alkanes comprise between 7-15% of the molecular components of natural terrestrial petroleum and are observed in petroleum synthesized by such as the Fischer-Tropsch processes. When these chiral molecules are created in low-pressure industrial processes, they occur always in equal enantiomer abundances, and the resulting synthetic petroleum does not manifest optical activity. In natural petroleum, these molecules occur in unequal enantiomer abundances, and the fluid manifests optical activity.

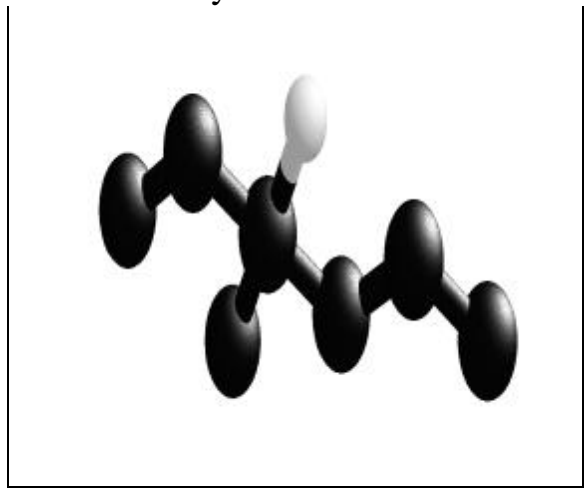

**Fig. 1 3-methylhexane, $C_7H_{16}$.**

**3. The explicit, general prediction of the genesis of optically active systems by classical thermodynamic argument.**

First is shown that the evolution of unbalanced abundances of enantiomers, scalemic mixtures, often results inevitably from the general requirements of thermodynamic stability theory. In keeping with the traditions of classical thermodynamics, no assumptions are made about any detailed properties of the material which composes the fluid mixture, other than the most basic attributes of their chirality.

From the cross derivatives of the differential of the Gibbs free enthalpy, $G(p,T,\{n_j\})$,

$$dG = -SdT + Vdp + \sum_j \mathbf{m}_j dn_j \qquad (1)$$

the differential equation for the chemical potential, as a function of pressure, at constant temperature is given as:

$$\left(\frac{\partial \mathbf{m}_i}{\partial p}\right)_{T,\{n_j\}} = \left(\frac{\partial V}{\partial n_i}\right)_{T,p,\{n_{i\neq j}\}} = V_i. \qquad (2)$$

With inclusion of the Gibbs mixing factor $RT\ln x_i$, the chemical potential of the $i$-th species is given in terms of its partial volume as:

$$\mu_i = \mu_i^{\ominus} + RT \ln x_i + \left( \int_{p^{\ominus}}^{p} V_i \, dp \right)_T. \tag{3}$$

The volumetric behavior of a multicomponent system can be described by the intensive variable, excess volume: $V^E = V - \Sigma_{(j)} x_j V_{m,j}$. When the formalism developed by Guggenheim[18] and Scatchard[19] is applied, the excess volume can be expressed as a series expansion:

$$V^E = \sum_i x_i \sum_{j<i} x_j \left\{ V_{0,ij}^E + V_{1,ij}^E (x_i - x_j) + V_{2,ij}^E (x_i - x_j)^2 + \cdots + V_{n,ij}^E (x_i - x_j)^n \right\}. \tag{4}$$

For the present analysis, the Guggenheim-Scatchard expansion, (4), may be truncated at the first term without loss of generality; when such is done, the molar volume may be written as:

$$V_m = \sum_j x_j V_{m,j} + \sum_j \sum_{k<j} x_j x_k 4 \left( V_{jk} \right)_{max}^E, \tag{5}$$

which approximation constitutes application of the Porter Ansatz. (The excess molar volumes in equation (5), $4 \left( V_{j,k} \right)_{max}^E$, contain the factor four in order that the it will return its maximum value in the case of a binary equimolar mixture at the racemic molar fractions $x_i = x_j$.)

### 3.1 The thermodynamic chirality function.

A thermodynamic system capable of manifesting optical activity may be considered generally as consisting of the two chiral enantiomers, L and D, together with a third achiral component, A. For such a three-component system, the chemical potential of the D-component enantiomer is:

$$\mu_D = \mu_D^{\ominus} + RT \ln x_D + \left( \int_{p^{\ominus}}^{p} V_D \, dp \right)_T. \tag{6}$$

When equation (5) is applied, together with the property that $\left( V_{DA} \right)_{max}^E = \left( V_{LA} \right)_{max}^E$, the partial molar volume of the D-component enantiomer becomes,

$$V_D = V_{mD} + x_D (1 - x_L) 4 \left( V_{DL} \right)_{max}^E + x_A^2 4 \left( V_{DA} \right)_{max}^E \; ; \tag{7}$$

and its chemical potential is thereby,

$$\mu_D = \mu_D^{\ominus} + RT \ln x_D +$$
$$+ \left( \int_{p^{\varnothing}}^{p} \left[ V_{mD} + x_L (1 - x_D) 4 \left( V_{DL} \right)_{max}^E + 4 x_A^2 \left( V_{DA} \right)_{max}^E \right] dp \right)_T. \tag{8}$$

Let the thermodynamic chirality functional, $Q(p,T;\vec{x})_{DL}$ be defined as the integral of the excess volume of the system of chiral molecules over its pressure as:

$$Q_{ij}(p,T;\vec{x}) = \frac{1}{RT}\left(\int_{p^{\ominus}}^{p} 4\left(V_{ij}(p,T;\vec{x})\right)^{E}_{max} dp\right)_{T}. \tag{9}$$

The chemical potential of the D-component enantiomer, (6) and (8), is then written:

$$\mathbf{m}_D = \mathbf{m}_D^{\ominus} + RT \ln x_D + \left(\int_{p^{\ominus}}^{p} V_{m,D} dp\right)_{T} + RT x_L (1-x_D) Q_{DL} + RT x_A^2 Q_{D,A}; \tag{10}$$

an analogous expression for $\mathbf{m}_L$, with the subscript L replacing that of D in equation (10).

### 3.2 Enantiomeric separation and conversion processes.

Three thermodynamic conditions characterize enantiomers: the equality of their reference chemical potentials; the equality of their molar volumes; and a (*usually*) non-vanishing excess volume of their mixtures:

$$\left.\begin{array}{l} \mathbf{m}_L^{\ominus} = \mathbf{m}_D^{\ominus} \\ V_{m,L} = V_{m,D} \\ V_{L,D}^{E} \neq 0 \end{array}\right\} \tag{11}$$

[The validity of the first two of equations (11) is intuitively obvious. The third of equations (11) may be considered to be (at this point) an assertion; it is proven in the following sections to hold usually.] The simultaneous requirements set forth by equations (11) are strictly thermodynamic ones and involve no detailed properties of the molecules themselves. In addition to the equalities of equations (11), the excess volume for either enantiomer with a non-chiral component, A, is identical, such that $Q_{DA} = Q_{LA}$.

Let it be assumed that the system is a reactive one, for the enantiomers can convert into one another, $D \rightleftarrows L$. The condition of chemical equilibrium, $\mathbf{m}_D = \mathbf{m}_L$, together with equations (10) and (11) and the equality $Q_{DA} = Q_{LA}$, gives:

$$\ln\frac{x_D}{x_L} - (x_D - x_L)Q_{D,L} = 0. \tag{12}$$

Equation (12) expresses succinctly and rigorously the essential determining factors for the spontaneous, abiotic, evolution of optical activity in fluids: the competition between the entropy of mixing, given by the first logarithmic term, and the oppositely-directed effects of the inevitable packing inefficiency of mixed enantiomers, given by the thermodynamic chirality function, $Q_{DL}$. Equation (12) defines the ana-lytic Lambert W function,[20, 21] which has always one real root and, as shown in Fig. 2., sometimes three.

The molar excess volume of a simple mixture often is relatively small, $V^E \sim$ 1-10 cm$^3$/mol.[22, 23] Therefore, at or near a pressure of one bar, for a system consisting only of the two enantiomers, $Q_{DL} = (1/RT)pV^E \approx (10^{-3}/RT)$ kJ. The denominator $RT$ is approximately 2.5 kJ at 300 K. Therefore, at modest pressures and essentially all temperatures, the second term in equation (12) cannot balance the first except at the value $x_D = x_L$, which solution describes the racemic mixture. Thus equation (12) establishes that, at low pressures, an abiotic system will usually evolve into a racemic mixture of enantiomers.

However, the thermodynamic chirality functional, $Q_{DL}$, depends directly upon the system pressure, for $Q_{DL} \sim V^E p$. For any non-vanishing, positive excess volume, $V_{DL}^E$, the thermodynamic chirality function has no limit as pressure increases. Therefore, for a system whose thermodynamic chi-rality function is *greater* than a certain threshold value, $(Q_{DL})^{threshold}$, there will always be a (usually, high) pressure above which the system will evolve *un*balanced abundances of enantiomers, and the resulting system will inevitably be optically active. For a system whose thermodynamic chirality function is *less* than the threshold value, $(Q_{DL})^{threshold}$, there will be no transition pressure, and such system will remain racemic. This behavior is shown graphically in Fig. 2 where are represented the plots of the two functions $\ln(x_D/(1-x_D))$ and $(2x_D-1)Q_{DL}$ (which correspond to a binary system composed solely of two enantiomers). As seen clearly in Fig. 2, for values of the thermodynamic chirality function, $Q_{DL}$, less than a threshold value, the system cannot evolve an unbalanced system. For the value, $Q_{DL} = 1.5$, there is only one solution, the racemic root, $x_D = 0.5$; for the value, $Q_{DL} = 4$, there is a second solution at the scalemic value, $x_D \approx 0.97$ (and a third, symmetrically, at $x_D \approx 0.03$).

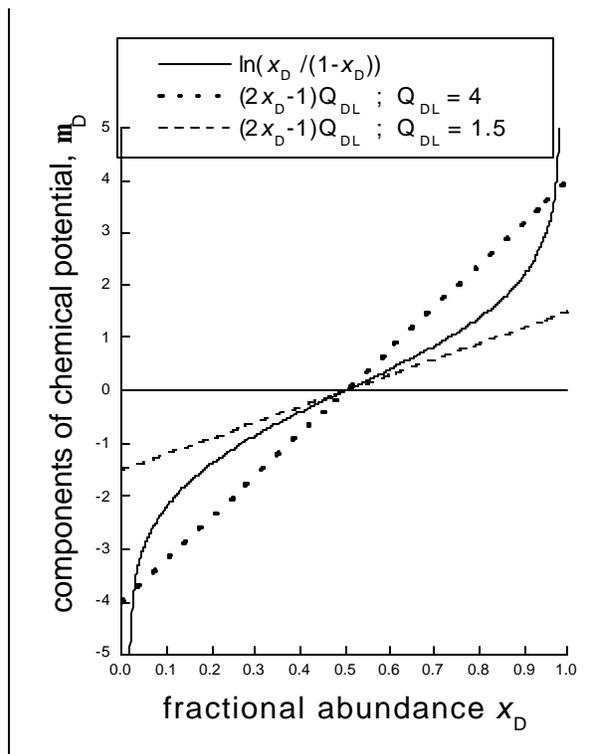

**Fig. 2 Values of the functions $\ln(x_D/(1-x_D))$ and $(2x_D-1)Q_{DL}$ which satisfy equation (12) at different values of the fractional abundance, $x_D$, in a binary mixture. Note the one racemic and two scalemic solutions to equation (12) for $Q_{DL} = 4$.**

For a general system composed of two enantiomers and a third component, A, the threshold value of $Q_{DL}$ for the onset of a racemic-scalemic transition is that for which the two terms in equation (12) have equal derivatives with respect to $x_D$; such that,

$$\frac{2}{(1-x_A)} = Q_{LD}. \qquad (13)$$

The foregoing thermodynamic argument has shown that chiral molecules in an unbalanced, scalemic, distribution of enantiomers can possess lower chemical potentials, and thereby lower Gibbs free enthalpy, than a racemic distribution of the same compound, under certain conditions of density. However, that argument gave no indication how a given distribution of enantiomers might, in new conditions of temperature or pressure, convert into a different one, such that the system could assume a lower free enthalpy.

For a chemically-reactive system, in which the achiral constituent molecules A and B are present, and in which the chiral molecules C evolve as,

$$A + B \longrightarrow C, \qquad (14)$$

the system will always transform into the distribution of reagents and products which possesses the lowest possible Gibbs free enthalpy. A chemically-reactive system will similarly evolve always that distribution of enantiomers which renders it the lowest free enthalpy.

A system which is not considered chemically-reactive, composed solely of the enantiomers $C_L$ and $C_D$ will also transform itself always into that distribution of enantiomers which renders it the lowest Gibbs free enthalpy. As a simple example, if either $C_D$ or $C_L$ enantiomer might result from a single-step reaction such as (14), the principle of detailed balance requires that there exists at equilibrium always a distribution of components,

$$A + B \rightleftharpoons \begin{cases} C_D \\ C_L \end{cases}, \qquad (15)$$

of which the abundances are determined by the general law of mass action. With a change of temperature or pressure, the equilibrium coefficient will change; and the abundance distribution of enantiomers will also change, so as to conform to the law of mass action and effect the minimum Gibbs free enthalpy. The constant, simultaneous production of enantiomers C and dissociation into constituent reagents A and B, in accordance with (15) and the principle of detailed balance, assure that the system always will evolve into the distribution of lowest free enthalpy.

### 3.3 The enantiomeric phase separation in scalemic mixtures.

If $Q_{DL}$ is positive and of sufficient magnitude, there is a threshold pressure at which the racemic mixture becomes thermodynamically unstable. If a conversion reaction D $\rightleftarrows$ L exists, the slightest excess of one enantiomer can make the system to develop a macroscopic excess of this enantiomer, when the pressure is raised further. Alternatively, the system can undergo a phase split into two scalemic phases. In either case, application of sufficient pressure always leads to the formation of phases with enantiomeric excess.

The molar Gibbs free enthalpy of the system of chiral enantiomers L and D together with an achiral component A is:

$$G_m = \sum_j x_j \boldsymbol{m}_j^{\ominus} + RT \sum_j x_j \ln x_j + \sum_j x_j \int_{p^{\ominus}}^{p} V_{m,j} dp + RT x_L x_D Q_{LD} + RT x_A (1-x_A) Q_{AD} \quad (16)$$

The limits of stability and critical points of mixtures are determined by the higher derivatives of the molar Gibbs free enthalpy, $G_{n,i} = (\partial^n G_m / \partial x_i^n)_{j,p,T}$, which gives for the D-enantiomer,

$$G_D = \left( \frac{\partial G_m}{\partial x_D} \right)_{x_L T, p} = \boldsymbol{m}_D^{\ominus} - \boldsymbol{m}_A^{\ominus} + RT(\ln x_D - \ln x_A) + \int_{p^{\ominus}}^{p} (V_{mD} - V_{mA}) dp + RT x_L Q_{LD} + RT(2x_A - 1) Q_{AD} \quad (17)$$

The second derivatives of $G_m$ are, respectively:

$$\left. \begin{array}{l} G_{2D} = \left( \dfrac{\partial^2 G_m}{\partial x_D^2} \right)_{x_L T, p} = RT\left( \dfrac{1}{x_D} + \dfrac{1}{x_A} \right) - 2RT Q_{AD} \\[1em] G_{LD} = \left( \dfrac{\partial^2 G_m}{\partial x_A \partial x_D} \right)_{x_L T, p} = RT \dfrac{1}{x_A} + RT Q_{LD} - 2RT Q_{AD} \\[1em] G_{2L} = \left( \dfrac{\partial^2 G_m}{\partial x_L^2} \right)_{x_L T, p} = RT\left( \dfrac{1}{x_L} + \dfrac{1}{x_A} \right) - 2RT Q_{AD} \end{array} \right\} \quad (18)$$

The conditions for phase stability require that:

$$G_{2D} G_{2L} - (G_{DL})^2 \geq 0 \quad (19)$$

The examination for phase separation, for which the equality holds in (19), introduces the equality

$$x_D = x_L = \frac{1}{2}(1 - x_A). \tag{20}$$

The roots of the equality (19) then admitted are:

$$Q_{DL} = \begin{cases} \dfrac{2}{(1-x_A)} \\ 4Q_{AD} - \dfrac{2}{x_A(1-x_A)} \end{cases}. \tag{21}$$

The first root of equation (21), for the general case of a three component system, corresponds to that determined by equation (12) for a binary system.

The molar Gibbs free enthalpy of a system containing two enantiomers and a third achiral component, as given by equation (16), has been calculated using the value for $Q_{DL}$ determined by the roots given by equation (21), and is shown in Fig. 3 as a function of the reduced molar fraction, $(x_D)_r$ which is the D fraction of the total molar fraction of the enantiomers, $(1-x_A)$.  (The reduced molar fraction has been used in order that the racemic solution will fall at the value 0.5.)  The double minima of the Gibbs free enthalpy as a function of $(x_D)_r$ is immediately apparent in Fig. 3, as is the fact that the racemic mixture, for those values of $Q_{DL}$, is at a maximum. That the two scalemic minima lie at equal values of Gibbs free enthalpy is shown by their double tangent.

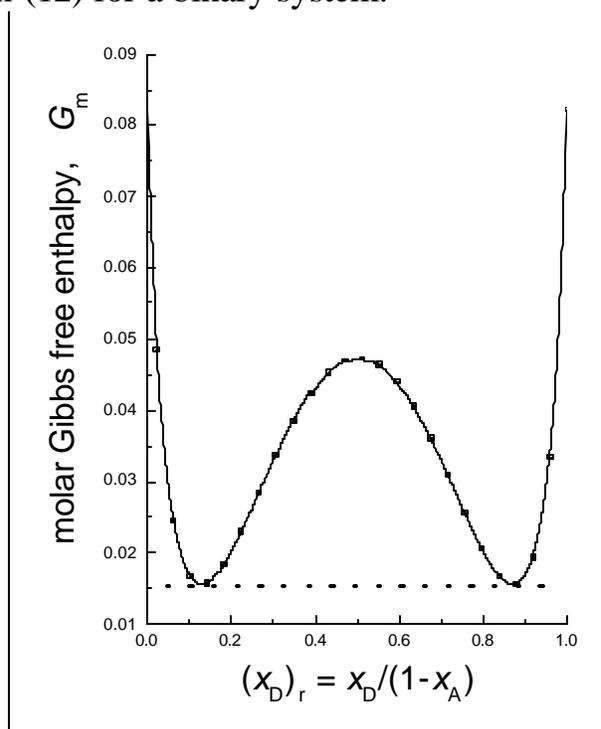

**Fig. 3 Gibbs free enthalpy at the roots of equation (21), (in arbitrary units).**

The foregoing analysis has involved determination of the equilibrium states of a multicomponent system, which contains a variable distribution of enantiomers together with a third achiral component, without considering the dynamic processes by which that system resolves to equilibrium when the pressure has changed. The two minima shown in Fig. 3 can be reached either by chemical reaction or by phase separation.

If the transition time for $L \to D$ conversion of the individual molecules is much slower than the transport diffusion time, then an initially racemic system will undergo a rapid physical separation, similar to gas-gas demixing;  and the system will resolve into two physically separate regions, one enriched in the D enantiomer, the other in the L. At the opposite extreme, if the transition time for L → D conversion is much faster than the diffusion time, the system will undergo a quasi-chemical type of reaction and change its constituent composition entirely. For cases in between, the system must be expected to undergo often complex chemical and dynamical behavior as it proceeds to equilibrium.

**4. Statistical mechanical calculation of the excess volume using the geometric properties of individual molecules.**

The thermodynamic analysis of the previous section established that a system of chiral particles will possess, in certain conditions of pressure, temperature, and degree of chirality, a lower free energy in a scalemic distribution than a racemic one. Consistent with the traditional perspective of classical thermodynamics, that analysis invoked no detailed properties of the chiral system, beyond the minimum functional definition of chiral enantiomers set out in equations (11). The thermodynamic analysis identified the system's excess volume, acting through the defined thermodynamic chirality parameter, as the operative physical property responsible for the racemic-scalemic transition. However, the classical thermodynamic analysis did not address the question how, or why, a system of chiral molecules ought, or must, manifest a non-vanishing excess volume. Nor did the thermodynamic analysis give any specific indication how the distribution of the chiral enantiomers might be determined from properties of those molecules.

In this section, the formalism is developed for direct calculation of the distribution of the chiral particles from the stereochemical properties of the individual molecules. First, a precise description of the geometry of chiral, convex, hard-body particles is developed by extension of the Kihara-Steiner equations. That geometric description is then used in the Pavlíček-Nezbeda-Boublík equations for convex, hard-body systems, from which an explicit expression is developed for the Helmholtz free energy of a multicomponent system which contains chiral particles.

**4.1. The geometric description of a hard-body system: Extension of the Kihara-Steiner equations.**

Convex hard-body systems have been described by Kihara and Steiner in terms of three geometric parameters, $\tilde{R}_i$, $\tilde{S}_i$, and $\tilde{V}_i$, which are determined by the support function that describes the volume and surface generated by the rolling of one hard body around the surface of another, in all possible orientations. As shown in the following subsection, the weighted products of these geometric entities, $\tilde{R}_i$, $\tilde{S}_i$, and $\tilde{V}_i$, enter the equation of state for systems of convex hard bodies and determine its thermodynamic properties. The parameter functional $\tilde{R}_i$ represents the mine its thermodynamic properties. The parameter functional $\tilde{R}_i$ represents the averaged radius of curvature of the support function and $\tilde{S}_i$ its averaged surface area; in this instance, $\tilde{V}_i$ does *not* represent the *i*-th partial volume but the effective volume determined by the support function. The individual functionals $\tilde{R}_i$, $\tilde{S}_i$, and $\tilde{V}_i$ are defined by the equations:

$$\left. \begin{aligned} \tilde{R}_i &= \frac{1}{4\pi} \int_0^\pi \int_0^{2\pi} \vec{r}_i \cdot \left( \frac{\partial \vec{u}_i}{\partial \theta} \times \frac{\partial \vec{u}_i}{\partial \phi} \right) d\theta\, d\phi \\ \tilde{S}_i &= \int_0^\pi \int_0^{2\pi} \vec{u}_i \cdot \left( \frac{\partial \vec{r}_i}{\partial \theta} \times \frac{\partial \vec{r}_i}{\partial \phi} \right) d\theta\, d\phi \\ \tilde{V}_i &= \frac{1}{3} \int_0^\pi \int_0^{2\pi} \vec{r}_i \cdot \left( \frac{\partial \vec{r}_i}{\partial \theta} \times \frac{\partial \vec{r}_i}{\partial \phi} \right) d\theta\, d\phi \end{aligned} \right\} \quad (22)$$

In equations (22), $\vec{u}(\theta,\phi)$ is the unit vector in the direction of the normal of the supporting plane, and $\vec{r}(\theta,\phi)$ is the vector from the origin to the contact point of the convex body with the supporting plane, as indicated in Fig. 4; the angles $\theta$ and $\phi$ are polar angles.

In his derivation of the functionals $\tilde{R}_i$, $\tilde{S}_i$, and $\tilde{V}_i$, Kihara assumed that the convex hard body possesses a center of inversion, and used such property to simplify his equations. However, a convex hard body need not be invariant under inversion. In order to represent a system whose components possess chirality, an appropriate tensorial description must be used. The expressions in the integrands of equations (22) are tensor entities, specifically pseudo-scalars; strictly each should be expressed similarly as:

$$\tilde{R}_i = \frac{1}{4p} \cdot \frac{1}{3} \cdot \left\{ \left( \begin{bmatrix} \int_0^p \int_0^{2p} \vec{r}(q,f)_i \cdot \left( \frac{\partial \vec{u}_i(q,f)}{\partial q} \times \frac{\partial \vec{u}_i(q,f)}{\partial f} \right) dq df + \\ \frac{1}{2} \begin{bmatrix} + \left[ \int_0^p \int_0^{2p} \vec{r}(q,f)_i \cdot \left( \frac{\partial \vec{u}_i(q,f)}{\partial q} \times \frac{\partial \vec{u}_i(q,f)}{\partial f} \right) dq df \right]_{\substack{x \to -x \\ y = y \\ z = z}} \end{bmatrix} \end{bmatrix} + \begin{bmatrix} \int_0^p \int_0^{2p} \vec{r}(q,f)_i \cdot \left( \frac{\partial \vec{u}_i(q,f)}{\partial q} \times \frac{\partial \vec{u}_i(q,f)}{\partial f} \right) dq df + \\ +\frac{1}{2} \begin{bmatrix} - \left[ \int_0^p \int_0^{2p} \vec{r}(q,f)_i \cdot \left( \frac{\partial \vec{u}_i(q,f)}{\partial q} \times \frac{\partial \vec{u}_i(q,f)}{\partial f} \right) dq df \right]_{\substack{x \to -x \\ y = y \\ z = z}} \end{bmatrix} \end{bmatrix} \right) \\ + \text{ similar terms in } \begin{Bmatrix} x = x \\ y \to -y \\ z = z \end{Bmatrix} \& \begin{Bmatrix} x = x \\ y = y \\ z \to -z \end{Bmatrix} \right\}$$

(23)

Clearly, in cases for which the hard body possesses a center of inversion, the terms in the second set of square brackets in equation (23) cancel, and the terms in the first set of square brackets are identical. Equation (23), and the analogous equations involving the Kihara surface and volume functionals, admit the representation of the geometric functionals, $\tilde{R}_i$, $\tilde{S}_i$, and $\tilde{V}_i$, as first-rank tensors:

$$\left. \begin{aligned} \vec{\tilde{R}}_i &= \tilde{R}_i^S \vec{e}_S + \tilde{R}_i^A \vec{e}_A \\ \vec{\tilde{S}}_i &= \tilde{S}_i^S \vec{e}_S + \tilde{S}_i^A \vec{e}_A \\ \vec{\tilde{V}}_i &= \tilde{V}_i^S \vec{e}_S + \tilde{V}_i^A \vec{e}_A \end{aligned} \right\},$$

(24)

in which the two orthogonal unit vectors, $\vec{e}_S$ and $\vec{e}_A$, define the two-dimensional vector space for the symmetric and anti-symmetric components, respectively, of the geometric functionals. In the first of equations (24), $\tilde{R}_i^S$ represents the symmetric terms in the first set of square brackets of equation (23), and $\tilde{R}_i^A$ the anti-symmetric difference terms in the second set; there are analogous Kihara functionals $\tilde{S}_i$, and $\tilde{V}_i$, in the second and third respectively. The symmetric contributions in (24) are always positive; the asymmetric contributions can be either positive or negative, depending upon the particular geometry of the body.

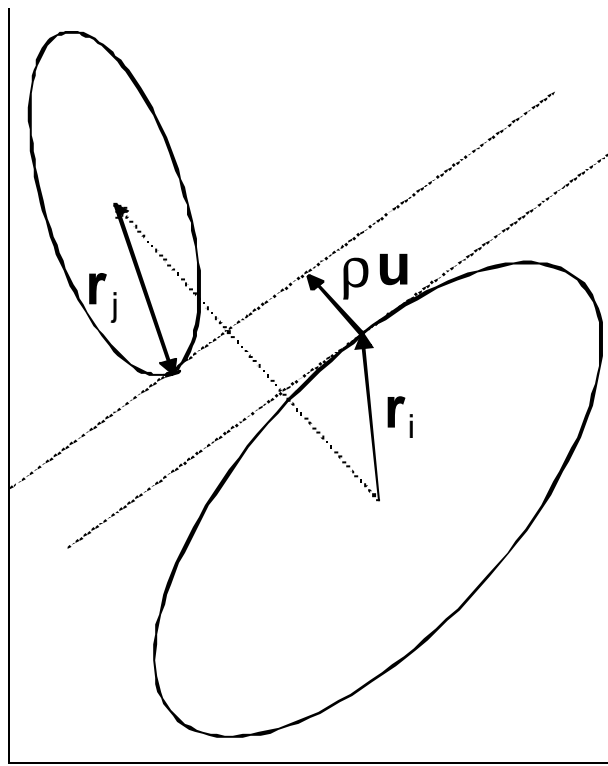

**Fig. 4 The geometric parameters used by Kihara to determine *R*, *S*, and *V* in equations (22).**

For example, consider again a tetrahedron, which is a convex body. Imagine an initially regular tetrahedron, with the apices of its base at [(a,0,0), (-a,0,0,), (0,√3/2a,0)], and its resulting geometry after that base has been deformed (with the fourth apex, out of the plane, remaining fixed) such that the base apices are at the new positions [(a,0,0), (-a,0,0,), (***d***,√3/2a,0)], - i.e., by a deformation by which one of the apices is moved parallel to its opposite side. The hard body which results from such a distortion, while still convex, is chiral; and a collection of molecules of such geometry will be optically active. The same tetrahedron similarly but oppositely deformed, with the position of its third apex above the plane of its base remaining fixed in both cases, to positions [(a,0,0), (-a,0,0,), (-***d***,√3/2a,0)], is also chiral, but clearly of opposite chiral sense. Clearly, a mixture of regular tetrahedra and ones deformed as described above cannot occupy the same minimum volume as would the regular tetrahedra alone, not even if the height of the apex above the deformed base were decreased so as to maintain the same hard-body volume. Equally clearly, the product of the radius of curvature of the deformed tetrahedron and the surface of the regular one would be increased, and, to terms linear in the deformation, approximately by a factor proportional to ***d***. It deserves to be noted that the parame-ter of distortion, *d*, need *not* be necessarily small; the base of the tetrahedron can be made very scalene, and the tetrahedron will remain a convex hard body.

### 4.2. Extension of the Pavlícek-Nezbeda-Boublík equations for mixtures of aspherical hard-body systems.

In order to develop the most general analysis possible of the thermodynamic stability of systems of chiral particles, such are described by the statistical mechanical formalism of hard bodies. The equation of state for the pure hard-sphere gas given by scaled particle theory (SPT)[24] represents one of the (very) few exactly-solvable problems in modern statistical mechanics. Significantly, scaled particle theory developed its major component, - the equation for the probability that the center of a particle exists within a radial shell at a certain distance, subject to the condition that the region defined by such radius be devoid of particles, - from analysis of statistical geometry. Later, Reiss, Frisch, and Lebowitz [25] extended SPT to describe mixtures of hard spheres. Twenty years after its first enunciation, scaled particle theory was extended further by Pavlícek, Nezbeda, and Boublík to describe mixtures of convex, hard-body gases,[26] using essentially arguments from statistical geometry and the precise descriptive parameters for such developed by Kihara and Steiner. Because the Pavlícek-Nezbeda-Boublík equations represent an extension of scaled particle theory, they share its rigor and precision, as well as the same of the differential geometric formalism of Steiner. For that reason, the Pavlícek-Nezbeda-Boublík equations have been used to examine the statistical thermodynamic stability of mixtures of chiral hard bodies.

The Pavlícek-Nezbeda-Boublík equation of state is:

$$\frac{p}{RT} = r\left[1 + \left(\frac{h}{1-h} + \frac{(\tilde{r}\tilde{s})}{r(1-h)^2} + \frac{h(12\tilde{t} - 5\tilde{s})\tilde{q}\tilde{s}}{9r(1-h)^3} + \frac{3r\tilde{q}\tilde{s}^2 - 2(\tilde{r}\tilde{s})^2 + 12(\tilde{r}\tilde{s})(\tilde{r}\tilde{t}) - 18(\tilde{r}\tilde{s})(\tilde{w}\tilde{t})}{9r^2(1-h)^3}\right)\right] \quad (25)$$

in which the molecular geometric functionals were originally defined by Pavlícek, Nezbeda, and Boublík as:

$$\left.\begin{array}{l}\tilde{q} = rq = r\sum_i x_i \tilde{R}_i^2 \quad \tilde{r} = rr = r\sum_i x_i \tilde{R}_i \quad \tilde{s} = rs = r\sum_i x_i \tilde{S}_i \\ \tilde{t} = rt = r\sum_i x_i \frac{\tilde{V}_i}{\tilde{R}_i} \quad \tilde{w} = rw = r\sum_i x_i \frac{\tilde{V}_i}{\tilde{S}_i} \quad h = r\sum_i x_i \tilde{V}_i = ru\end{array}\right\} . \quad (26)$$

The shape-dependent parameters *q*, *r*, *s*, *t*, and *w* are weighted sums of the functionals of the respective support functions which describe the volume and surface generated by the rolling of one hard body around the surface of another, in all possible orientations; **h** represents the weighted packing fraction, $x_i$ the molar fractions.

orientations; $\mathbf{h}$ represents the weighted packing fraction, $x_i$ the molar fractions. The individual functionals $\tilde{R}_i$, $\tilde{S}_i$, and $\tilde{V}_i$ are defined tensorially by the equations (24). The tensor contraction of their products, such as $\tilde{r}\tilde{s}, \tilde{q}\tilde{s}, \tilde{r}\tilde{t}$, etc., which enter the Pavlícek-Nezbeda-Boublík equation of state, (25), are defined using the two-dimensional identity matrix and the first of the Pauli matrices,

$$\left. \begin{array}{l} \begin{pmatrix} 1 & 0 \\ 0 & 1 \end{pmatrix} = \mathbf{I}_2 \\ \begin{pmatrix} 0 & 1 \\ 1 & 0 \end{pmatrix} = \mathbf{s}_1^{Pauli} \end{array} \right\}. \tag{27}$$

Because the vector space defining the degree of geometric chirality spans two dimensions, only one of the Pauli matrices are required; and the tensor properties of the geometric functionals are described using the symmetric and anti-symmetric operators

$$\left. \begin{array}{l} \mathbf{s}^S = \mathbf{I}_2 + \mathbf{s}_1^{Pauli} \\ \mathbf{s}^A = 2\mathbf{s}_1^{Pauli} \end{array} \right\}. \tag{28}$$

The contraction of the geometric vectors, $\mathbf{v}_1$ and $\mathbf{v}_2$ is then:

$$v_1 v_2 = \mathbf{v}_1 \wedge \mathbf{v}_2 = \mathbf{v}_1 \left( \mathbf{s}^{Chiral} \right) \mathbf{v}_2, \tag{29}$$

where $\mathbf{s}^{Chiral} = \mathbf{s}^S + \mathbf{s}^A$. For example, the contraction of the geometric vectors $r$ and $s$ is thereby:

$$rs = \left( \sum_i x_i \vec{\tilde{R}}_i \right) \left( \mathbf{s}^{Chiral} \right) \left( \sum_j x_j \vec{\tilde{S}}_j \right). \tag{30}$$

For the case where there exist only the two enantiomers, D and L, the product function $rs$ given by equation (29) and (30) has the form:

$$\begin{aligned} (rs)_{DL} = & \left( x_D \vec{\tilde{R}}_D \right) \mathbf{s}^{Chiral} \left( x_L \vec{\tilde{S}}_L \right) = \\ = & x_D^2 \left( \tilde{R}_D^S \tilde{S}_D^S + \tilde{R}_D^A \tilde{S}_D^S + \tilde{R}_D^S \tilde{S}_D^A + \tilde{R}_D^A \tilde{S}_D^A \right) \\ & + x_D x_L \left( \tilde{R}_D^S \tilde{S}_L^S + \tilde{R}_L^S \tilde{S}_D^S + \tilde{R}_D^S \tilde{S}_L^A + \tilde{R}_L^A \tilde{S}_D^A \right) \\ & + x_L^2 \left( \tilde{R}_L^S \tilde{S}_L^S + \tilde{R}_L^A \tilde{S}_L^S + \tilde{R}_L^S \tilde{S}_L^A + \tilde{R}_L^A \tilde{S}_L^A \right) \end{aligned}. \tag{31}$$

Plainly, equations (29) and (31) return the identical expressions for symmetric, achiral molecules as do the original Kihara-Steiner equations. Equally plainly, equations (29) and (31) return the correct equations for single-component systems, given by either the first or third lines of the expansion of $(rs)_{DL}$. Most importantly, the second line of the expansion of $(rs)_{DL}$ gives an additional contribution *only* in mix-tures. There are similar contributions for chirality for the other functionals which appear in the Pavlícek-Nezbeda-Boublík equation of state, $(\tilde{r}\tilde{t})_{DL}$, $(\tilde{w}\tilde{t})_{DL}$, $(\tilde{q}\tilde{s})_{DL}$, etc. Such cross-terms, which appear only in mixtures, have all the units of volume and are responsible for the excess volume. For two enantiomers, the anti-symmetric components possess identical magnitude and opposite sign:

$$\left.\begin{array}{l} \vec{\tilde{R}}_D = \tilde{R}_D^S \vec{e}_S + \tilde{R}_D^A \vec{e}_A \\ \vec{\tilde{R}}_L = \tilde{R}_D^S \vec{e}_S - \tilde{R}_D^A \vec{e}_A \end{array}\right\}, \tag{32}$$

and similarly for both components of $\tilde{S}_D$ and $\tilde{V}_D$ for both enantiomers L and D. When equation (32) is applied, the "excess" parameter $(rs)^E$ (the second line of the expansion of $(rs)_{DL}$) takes on the particularly simple form:

$$\left((rs)_{DL}\right)^E = x_D x_L \left(\tilde{R}_D^S \tilde{S}_D^S - \tilde{R}_D^A \tilde{S}_D^A\right). \tag{33}$$

Defining the terms in the first (or third) line of the expansion of $(rs)_{DL}$ as $(rs)_{DL}^0$ the product-functional $rs$ may be written for a binary system as

$$(rs)_{DL} = (1 - 2x_D + 2x_D^2) rs_{DL}^0 + 4x_D(1 - x_D) rs_{DL}^E. \tag{34}$$

For a racemic mixture, the product-functional $(rs)_{DL}$ has the especially simple form

$$\left((rs)_{DL}\right)^{racemic} = rs_{DL}^0 + rs_{DL}^E = rs_{DL}^0 \left(1 + e_{rs}^{Chr}\right). \tag{35}$$

There are similar expressions for the other product-functionals, $rt$, $ts$, etc., which will involve similar chirality parameters, $e_{rt}^{Chr}, e_{ts}^{Chr}$, etc. In the following analysis, a "minimalist" perspective is taken, such that the effects of chirality are taken into account for only the product-functional $rs$; therefore the subscripts will be dropped from the parameter of chirality, $e_{rs}^{Chr}$.

Because this analysis involves the geometry properties of the individual molecules, the system density, $\rho$, has been factored out of the geometric functionals so as to return the alternate form of the Pavlícek-Nezbeda-Boublík equation of state:

$$\frac{(p)^{PNB}}{\rho k_B T} = 1 + \frac{\eta(A + B\eta + C\eta^2)}{(1 - \eta)^3} \tag{36}$$

in which

$$A = \left(1 + \frac{rs}{u}\right)$$

$$B = -\left(2 + \frac{(rs)}{u} + \frac{1}{u^2}\left(2(rs)tw - \frac{4r^2 st}{3} + \frac{\left(2(rs)^2 - 3qs^2\right)}{9}\right)\right) \quad (37)$$

$$C = \left(1 + \frac{1}{u^2}\left(\frac{4qst}{3} - \frac{5qs^2}{9}\right)\right)$$

When these molecular shape variables, $A$, $B$, and $C$ are used, the Helmholtz free energy of the mixed hard-body system is written as:

$$F = Nk_BT\left[-\ln\left(\frac{V}{N\mathbf{l}^3}\right) - 1 - C\ln(1-\mathbf{h}) + \frac{\mathbf{h}(D\mathbf{h}+E)}{(1-\mathbf{h})^2}\right] \quad (38)$$

in which $\mathbf{l}$ is the thermal de Broglie wavelength, and the variables $D$ and $E$ are related by,

$$\left.\begin{array}{l} D = \dfrac{3C + B - A}{2} \\ E = A - C \end{array}\right\}. \quad (39)$$

It should be noted that the functions $r$, $s$, $t$, $w$, and $\mathbf{u}$, and thereby also $A$, $B$, $C$, $D$, and $E$, depend upon the molecular geometric properties *only* and are not thermodynamic variables. Equations (36), (37), and (38) have been applied to calculate the Helmholtz free energies, Gibbs free enthalpies, and thermodynamic Affinities of a large number of racemic and single-component, chiral hard-body systems of diverse geometric shape and degree of chirality, over a wide range of pressure and temperature.

## 5. The racemic-scalemic transition: The onset of unequal abundances of enantiomers in a chiral system at high density.

The direction of spontaneous evolution of any system is determined by its thermodynamic Affinity, $A(p,T,\{n_j\})$. The second law of thermodynamics, expressed mathematically by De Donder's inequality,

$$\left.\begin{array}{l} dQ' = A d\mathbf{x} \geq 0 \\ A = -\sum_{i,r,a} \mathbf{n}_{i,r} \mathbf{m}^a_{i,r} \end{array}\right\}, \quad (40)$$

requires that the thermodynamic Affinity be positive for any spontaneous transition, for which the variable of extent is $\mathbf{x}$.[27-30]

The process of evolution here examined is that of the racemic-scalemic transition, which may be expressed as:

$$\frac{1}{2}C_D + \frac{1}{2}C_L \longrightarrow x_D C_D + (1-x_D)C_L, \qquad (41)$$

where, in the scalemic state (right side of (41)), the molar fraction $x_D \neq 1/2$. The thermodynamic Affinity for the onset of an unbalanced system of single-component enantiomers, from one initially racemic, is:

$$A(p,T;\mathbf{e}^{Chr},\mathbf{a})^{racemic \to pure} = \left(\frac{1}{2}m_D + \frac{1}{2}m_L\right)^{racemic} - (m_D)^{pure}. \qquad (42)$$

When the thermodynamic Affinity, (42), is negative, the system either evolves in the opposite direction, toward a racemic mixture, or remains as such. However, whenever this thermodynamic Affinity becomes positive, a racemic mixture transforms to a scalemic one with unequal abundances of enantiomers. For a single-mole system, the thermodynamic Affinity is simply the difference in the Gibbs free enthalpy of the racemic and pure systems.

### 5.1  Theoretical analysis of a chiral hard-body system.

The Gibbs free enthalpy has been calculated over the range of pressures 1-500 kbar, at temperatures of 300 K, and 1000 K, for a wide collection of convex hard-body systems of different molecular geometries, whose molecular mass and volume, $\mathbf{u}$, were taken to be, respectively, 105 gm/mol and 65 cm$^3$/mol, which values correspond roughly to those for single-branched heptane or octane. The thermodynamic Affinity was calculated for general convex hard-body systems characterized by values of asphericity $\mathbf{a} = rs/(3\mathbf{u}) = 1.5, 2.0, 2.5, 3.0, 4.0, 5.0$, and for values of the prolateness parameter, $\mathbf{b} = r^2/s = 1.75, 2.00, 2.25, 2.5, 3.0, 4.0$. Identical calculations were performed also for prolate ellipsoids with ratios of long to short axes of 1.75, 2.0, 2.5, 3.0, 4.0.

For each geometry, the respective system was examined for the following values of the chirality parameter: $\mathbf{e}^{Chr} = 0.025, 0.05, 0.1, 0.5, 1.0$. The Gibbs free enthalpy was calculated in each case for both the racemic and single-component systems. Because a typical experimental value for measured molar excess volume is approximately 1-10 cm$^3$/mol for simple systems,[23, 31] for several geometries studied, the chirality parameter, $\mathbf{e}^{Chr}$, was specifically chosen to produce an excess volume of 1 cm$^3$ at STP for a hard-body gas with the approximate molecular and thermodynamic characteristics of octane. The choice of 1 cm$^3$/mol for molar excess volume was chosen as most reasonably conservative.

Consistent with the intention to take a conservative, "minimalist" approach for the analysis, the effects of chirality were restricted to the product of the molecular parameters *r* and *s*; such that the product *rs* which appears in the Pavlícek-Nezbeda-Boublík equations was applied as given in equation (31). All other molecular shape variables, *q*, *s*, *t*, *w*, and their powers and products with *rs* have been left unchanged. Such restriction constitutes a formidable *under*estimation of the effects of chirality upon the thermodynamic functions, particularly upon the Gibbs free enthalpy. An effect of chirality is always to *increase* the effective asphericity of the hard body, and thereby also to increase the hard-body components of both the pressure and the Gibbs free enthalpy; an addition of chirality cannot ever decrease the magnitude of those functions, at any density or temperature. The property of aspherical hard-body systems has been demonstrated previously.[32]

    The results of all calculations are qualitatively identical. At low pressures, and thereby low densities, the Gibbs free enthalpy of the racemic mixture is always lower, at all temperatures and for every geometry. The contribution to the Gibbs free enthalpy attributable to the entropy of mixing dominates the system at low pressures, regardless of geometry. However, at high pressure and for all values of the chirality parameter greater than a certain threshold value, the contribution to the Gibbs free enthalpy attributable to its excess volume exceeds that of its entropy of mixing, and the Affinity abruptly changes sign, and remains positive for the racemic-scalemic transition.

An example of the effect of pressure upon the thermodynamic Affinity is shown in Fig. 5, on a logarithmic scale of pressure through the range 1-500 kbar, for typical systems of pure and racemic particles of molecular mass 105 g/mol and characteristic volume, $u = 65\text{ cm}^3$, with asphericity, $a = 1.5$, and degree of prolateness, $b = 2.25$, for values of its chirality parameter, $e^{Chr} = 0.025, 0.05, 0.1, 0.5, 1.0$. These parameters of molecular geometry represent very conservative measures of asphericity, prolateness, and chirality; the magnitude of $e^{Chr}$ is often substantially greater than one. The asphericity, $a = 1.5$, describes a molecule only modestly deformed from spherical, and degree of prolateness, $b = 2.25$, characterizes one for which the length of its major axis is slightly more than twice that of its minor axis; these parameters are comparable to the same, for example, of pro-pane or butane, although assigned to a larger molecule. The degrees of chirality, for which the greatest is $e^{Chr} = 1.0$, are also modestly taken. Mathematical modeling for molecules as simple as 3-methylhexane indicates that the difference in the product of the radius of curvature and the characteristic surface of one enantiomer, and of its mirror image, can easily be as great as 2-4. As shown in Fig. 5, an increase of pressure has no discernable effect upon the system below 1 kbar, excepting only for molecules of the greatest degrees of both asphericity and chirality. However, above 1 kbar, the product of the excess volume and pressure begins to shift the relative values of the Gibbs free enthalpy of, respectively, the racemic and scalemic system. At pressures higher than 10 kbar, the additional component to the Gibbs free enthalpy of the product of the excess volume and pressure begins to dominate the system; and, at sufficiently high pressure, typically on the order of 20,000-450,000 atm, the contribution to the Gibbs free enthalpy of the racemic mixture attributable to its excess volume becomes greater than that attributable to its entropy of mixing. At the density corresponding to such pressure, the thermodynamic Affinity of the system changes sign, and at greater densities, the Gibbs free enthalpy of the scalemic system is lower than that of the racemic one. As shown also in Fig. 5, for the lowest value of the parameter of chirality, $e^{Chr} = 0.025$, the thermodynamic Affinity does not change sign, and the system would remain racemic at all pressures, which behavior is consistent with the purely thermodynamic argument of section 2.

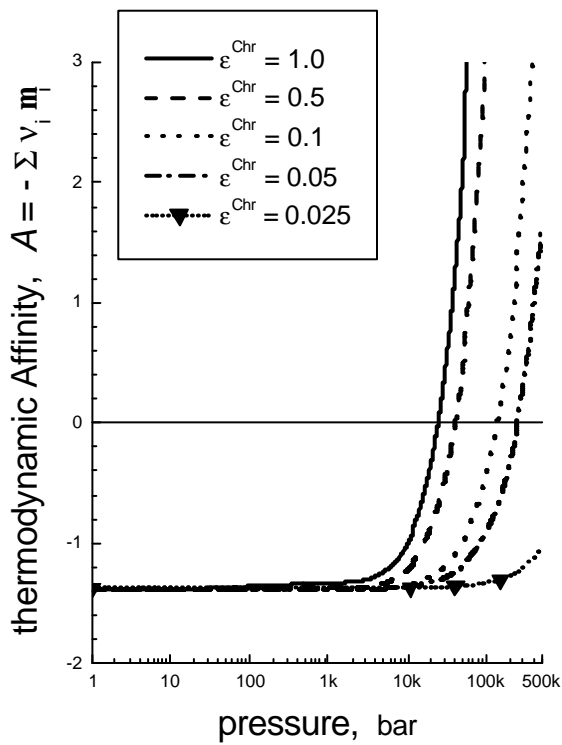

**Fig. 5 The thermodynamic Affinity, $A(p,T;a)$ of a chiral hard-body system of asphericity $a = 1.5$ and prolateness $b = 2.25$ as a function of pressure for different values of the parameter of chirality, $e^{Chr}$.**

In Fig. 6, are shown the thermodynamic Affinities for the representative hard-body system, for each of the test values of the parameter of chirality, on a linear pressure scale between 1-500 kbar. That the Affinity for the lowest value of the chiral parameter, $e^{Chr} = 0.025$, does not change sign shows clearly. Although the traces in Fig. 6 appear linear, they are not; the trace for the lowest value of the chiral parameter has a negative second derivative, and, at high pressures, the Gibbs free enthalpy of the single-component system increases less rapidly than that for the racemic one. The Affinity, for that and lesser values of $e^{Chr}$, never crosses the $x$-axis; and the system remains racemic.

If the molecular specie comprising the enantiomers is generated at high pressure, as for example, hydrocarbon molecules from hydrogen and carbon, the system will evolve, above the racemic-scalemic transition pressure, a strongly unequal distribution of enantiomers for which the abundance ratio will be determined by thermodynamic Affinity and the general law of mass action. If the system was initially composed of both enantiomers, above the racemic-scalemic transition pressure, it will either transform into one of an unequal distribution of enantiomers, or will undergo phase separation; in such case, its specific evolution will depend upon the temperature and the transition rate for the processes by which one enantiomer transforms into its chiral opposite.

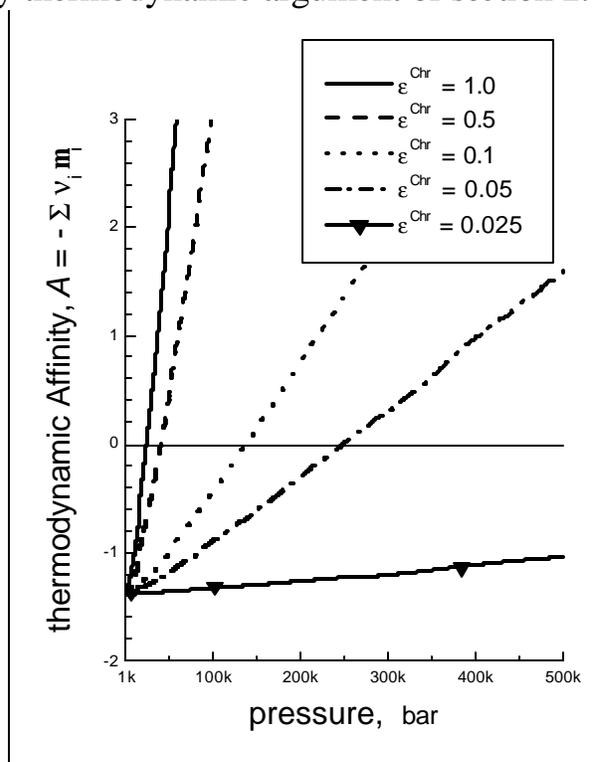

**Fig. 6 The thermodynamic Affinity, $A(p,T;\mathbf{a})$ of a chiral hard-body system of asphericity $\mathbf{a} = 1.5$ and prolateness $\mathbf{b} = 2.25$ as a function of pressure for different values of the parameter of chirality, $e^{Chr}$, on a linear scale.**

## 5-2. Observation of the evolution of the chiral molecules CHFClI and $C_8H_{12}$ by Monte Carlo simulation.

The previous theoretical analysis was restricted to convex, hard-body molecules. Such restriction allowed formal mathematical precision at a cost of generality. In order to investigate the racemic-scalemic transition of more realistic chiral molecules, isothermal-isobaric Monte Carlo simulations have been performed, over wide ranges of density, to calculate the excess volumes of two fluids consisting of the fused-hard-sphere molecules: fluorochloroiodomethane, CHFClI, and 4-vinylcyclohexene, $C_8H_{12}$. Fluorochloroiodomethane is obviously both chiral and not convex, as can be seen in its "ball-&-stick" representation in Fig. 7. The molecule 4-vinylcyclohexene is strongly concave, as shown in its "stick diagram" representation in Fig. 8.

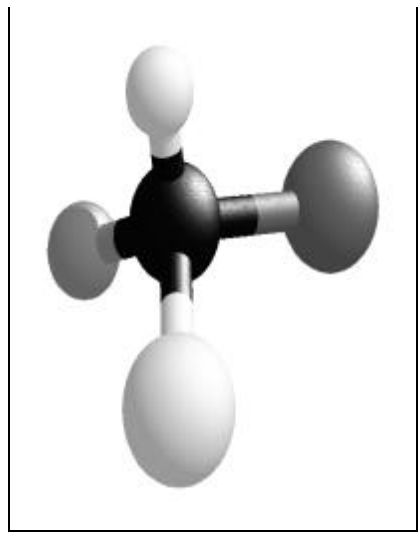

**Fig. 7 Fluorochloroiodomethane, CHFClI.**

The atomic coordinates for CHFClI were obtained with the MM3 computer packages of Allinger *et al.*,[33-35] and those of 4-vinylcyclohexene were known from previous work. For both molecules each atom was modeled as a hard sphere with the appropriate van der Waals radius, which were taken from the data compiled by Emsley[36] and reduced approximately 10% to agree better with experimental data of their fluid densities.

The simulation ensembles contained 256 or 512 molecules, either pure enantiomers or a racemic mixture. A simulation cycle consisted of attempted moves, - translation and rotation, - of all molecules, followed by a random volume change. After the system reached equilibrium, such that the volume remained constant throughout a cycle, 150000-2500000 cycles were executed to obtain the thermodynamic averages.

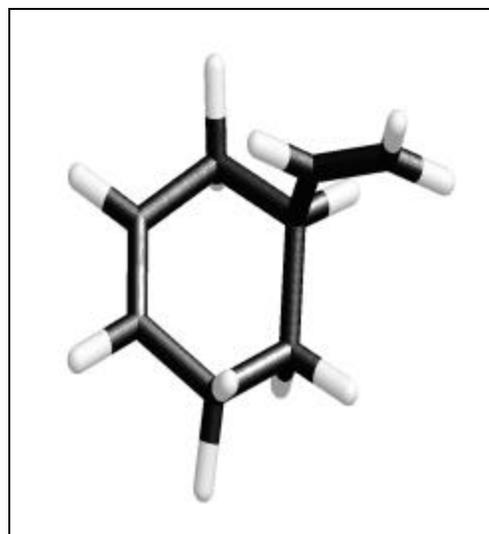

**Fig. 8 4-vinylcylohexene, $C_8H_{12}$.**

The fixed property in the simulation of a hard-body system is the dimensionless pressure/temperature ratio:

$$p_r = \frac{l_s^3}{k_B T} p \tag{43}$$

where $l_s$ is the length unit of the simulation, taken for these investigations as 1 Angstrom. The calculated property is the system density or the molar volume. Thus a value $p_r = 0.3$ corresponds roughly to 12 kbar at 300 K. The difference of the molar volumes of, respectively, the racemic mixture and the pure enantiomer gives directly the excess volume. The simulations reported here were limited to a maximum dimensionless pressure of 0.3. At higher pressures, these samples undergo partial solidification by the Alder-Wainwright transition. Therefore, above such pressure, no meaningful density averages can be calculated.

The results of the Monte Carlo simulation investigations of fluorochloroiodomethane and 4-vinylcyclohexene are shown in Fig. 9. The negative values and fluctuations of the excess volumes of both CHFClI and $C_8H_{12}$ at low pressures are not important; for the thermodynamic chirality functional, $Q^{Chr}$, involves the integral of the excess volume over pressure. Therefore, the high-pressure values of $V^E$ determine both $Q^{Chr}$ and the Gibbs free enthalpy. It can be seen that, for both molecules studied, the pure enantiomer fluids have greater density at high pressures. For interpreting the excess volumes, the absolute uncertainty of the molar volumes and of the excess volumes can be estimated as:

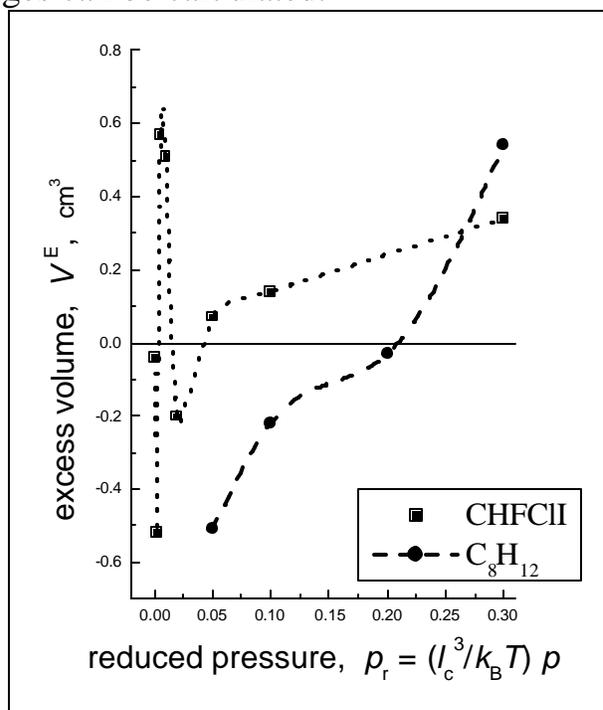

**Fig. 9 Excess volumes of fluorochloroiodomethane, CHFClI, and 4-vinylcyclohexene, $C_8H_{12}$, as functions of reduced pressure.**

$$\Delta V = \frac{dV}{dx_i} \Delta x_i = -\frac{V}{x_i} \Delta x_i \ . \tag{44}$$

Therefore, although the density of a dilute gas at the dimensionless pressure of 0.001 can be calculated quite accurately, the excess volume of the halogenated methane has (at that value of $p_r$) a statistical uncertainty of approximately 1 cm³/mol; and therefore, in that case, a value for the excess volume of 0.5 cm³/mol must be regarded as insignificant. However, at high pressures, the statistical uncertainty decreases to approximately $0.1 \text{ cm}^3/\text{mol}$. Therefore these calculated excess volumes must be regarded as significantly positive.

## 6. Discussion.

The phenomenon of optical activity in abiotic fluids has been shown in the previous sections to be a direct consequence of the chiral geometry of the system particles acting according to the laws of classical thermodynamics.

- In section 2, the purely thermodynamic argument was developed which relates the evolution of optical activity in a system of chiral molecules to the excess volume of scalemic mixtures. The thermodynamic analysis established that, above a threshold value of excess volume, the Gibbs free enthalpy becomes, at high densities, lower for an unequal, scalemic distribution of enantiomers.
- In section 3, the excess volume of a scalemic mixture of enantiomers has been related to their geometric properties using the Kihara-Steiner equations, which have been extended to describe particles which lack a center of inversion. The chiral property described by the extension of the Kihara-Steiner equations has been introduced into the Pavlícek-Nezbeda-Boublík equations for mixtures of hard bodies, with which are calculated the Gibbs free enthalpies and thermodynamic Affinities of the hard-body systems.
- In section 4, formal calculations of the thermodynamic Affinities show that, in accordance to dictates of the second law, a system of chiral molecules will often evolve unbalanced, scalemic, abundances of enantiomers at high densities. The results of Monte Carlo investigations on the chiral (and *non*-convex) molecules CHClI and $C_8H_{12}$ demonstrate positive, increasing excess volume at high pressures, the prerequisite for the racemic-scalemic transition.

Although the formal analysis in the preceding section considered model systems composed of hard bodies, the results there described hold true also, without qualification, for real molecules, - i.e., ones which possess additionally a long-range, attractive, van der Waals-type component to their intermolecular potential. The racemic-scalemic transition occurs at a density at which the free energy contribution attributable to the product of the system's excess volume and pressure exceeds that attributable to its entropy of mixing. The former is usually positive, the latter always negative. As indicated in equation (12) and Fig. 2, and as demonstrated using equations (36), (37), and (38), and as shown explicitly in Fig. 5, the onset of the racemic-scalemic transition occurs at increased densities. The dependence of the racemic-scalemic transition upon pressure results from its control of the density. The attractive, long-range, van der Waals-type component to the potential can only act to *in-* *crease* the density, at *all* temperatures. Therefore, the presence of a van der Waals-type component to the potential could not eliminate, or cause somehow to vanish, the onset of the racemic-scalemic transition in any fluid of real molecules the geometry of whose short-range, repulsive, - i.e., hard-body, - component of potential imposes an increase in the excess-volume component of the Gibbs free enthalpy the magnitude of which exceeds its mixing-entropy component.

Many of the qualitative features of the racemic-scalemic transition can be understood by considering the virial expansion of the chemical potentials which enter the thermodynamic Affinity. The chemical potential, when written in terms of the virial coefficients, has the form:

$$\boldsymbol{m}_i(p,T;\boldsymbol{a}) = \boldsymbol{m}_i^{I.G.}(p,T) + (B_2)_i p_i + \cdots, \tag{45}$$

in which $B_1$ represents the second virial coefficient in the density expansion of the compression factor $Z = pV/nRT$, and equals, to the first-order approximation of the virial expansion, the natural logarithm of the fugacity coefficient. For a mixed, hard-body gas, the second virial coefficient is, exactly,

$$B_2 = \boldsymbol{u} + rs = \left(\sum_i x_i \tilde{V}_i\right) + \left(\sum_i x_i \tilde{R}_i\right)\left(\sum_j x_j \tilde{S}_j\right); \tag{46}$$

and for the contribution of the $i$-th component,

$$(B_2)_i = \tilde{V}_i + \tilde{R}_i \tilde{S}_i + \tilde{R}_i\left(\sum_{j \neq i} x_j \tilde{S}_j\right) + \tilde{S}_i\left(\sum_{j \neq i} x_j \tilde{R}_j\right).$$
$$= \boldsymbol{u}_i^0 + (rs)_i^0 + (rs)_i^E \tag{47}$$

For a binary mixture of enantiomers, the excess contribution to the second virial coefficient is, when the approximation (24) is applied:

$$(rs)_{DL}^E = (rs)_{DL}^0 \boldsymbol{e}_{DL}^{Chr}. \tag{48}$$

Comparing equations (48) and (9), one may note that

$$Q_{L,D}(p,T) \sim 4rs\boldsymbol{e}_{DL}^{Chr}\frac{p}{RT}, \qquad \text{at low pressures.} \tag{49}$$

When the Kihara asphericity parameter, $\boldsymbol{a} = rs/3\boldsymbol{u}$, is introduced into equation (49), the thermodynamic chirality function may be approximated as,

$$Q_{L,D}(p,T) \sim 12\boldsymbol{uae}_{DL}^{Chr}\frac{p}{RT}. \tag{50}$$

The approximation (50) indicates qualitatively many of the principle characteristics of the racemic-scalemic transition, and, together with equations (12) and (13), provides a direct interpretation of the limitations of the thermodynamic chirality function described in section 2:

- For a combination of hard-body molecular volume, *u*, degree of asphericity, *a*, and parameter of molecular chirality, $e^{Chr}$, there will exist a (usually high) pressure at which the additional free energy attributable to the excess volume of a racemic mixture will exceed that attributable to its entropy of mixing.
- For a given parameter of chirality, $e^{Chr}$, the greater is the degree of asphericity, *a*, the lower will be the density and pressure of the racemic-scalemic transition; similarly, for a given degree of asphericity, *a*, the larger is the parameter of chirality, $e^{Chr}$, the lower will be the density and pressure of the racemic-scalemic transition.
- For any combination of the degree of asphericity, *a*, and the parameter of chirality, $e^{Chr}$, the larger is the hard-body molecular volume, *u*, the lower will be the density and pressure of the racemic-scalemic transition.
- If the product of the hard-body molecular volume, *u*, the degree of asphericity, *a*, and the parameter of chirality, $e^{Chr}$, is less than a threshold value, the system will never undergo the racemic-scalemic transition no matter how high its pressure.

As noted in the previous section, these easily-understood properties are exactly those demonstrated by many detailed calculations of binary chiral, hard-body systems of diverse degrees of asphericity, chirality, and particle volumes.

The phenomena of expansion upon mixing, and segregation (or phase separation) at high pressures, are interrelated. The pressure dependence of the complex phase behavior of binary liquid systems has been described generally by Rebelo.[37] The type of phase separation characteristic of the racemic-scalemic transition corresponds to liquid-liquid equilibrium [38] behavior described by Rebelo as either type 1 (or a) or type 3 (or c) as shown in Fig. 3 on page 4279 in the work cited. Both types of fluid behavior manifest phase separation with increased pressure and a a lower critical solution pressure transition [LCSP] point. Therefore, a mixture of enantiomers with favorable physical parameters should be expected to manifest an LCSP. When a liquid mixture of enantiomers coexists with its vapor under the constraints of equations (11), the system will inevitably present an azeotrope, for the equal pressures of both L and D enantiomers, combined with their non-ideal behavior, assures that the pressure must be at an extremum.

As noted in section 2, for a system to manifest optical activity, the onset of the racemic-scalemic transition must occur before the system undergoes the high-pressure, Alder-Wainwright, fluid-solid transition. The prediction of the Alder-Wainwright transition by scaled particle theory has already been discussed generally[39] and further investigated specifically for aspherical, hard-body systems.[32] The absolute limits of the fluid phase of any system are determined by the values of den-sity at which its entropy vanishes. The Pavlícek-Nezbeda-Boublík equations give such values by the roots of the equation:

$$\frac{5}{2} + \ln\left(\frac{V}{N\mathbf{l}^3}\right) = -\frac{(S^{hc})^{PNB}}{Nk_B}, \qquad (51)$$

in which the Pavlícek-Nezbeda-Boublík hard-core component of the entropy, $(S^{hc})^{PNB}$, is:

$$(S^{hc})^{PNB} = -Nk_B\left[C\ln(1-\mathbf{h}) + \frac{\mathbf{h}(D\mathbf{h}+E)}{(1-\mathbf{h})^2}\right]. \qquad (52)$$

The geometric functionals, $C$, $D$, and $E$, in equation (52) are defined in equations (37) and (39); and from those equations, one can perceive quickly that the complexity of the relationships between the Kihara-Steiner functionals $\tilde{R}_i$, $\tilde{S}_i$, and $\tilde{V}_i$ do not permit a simple estimate for predicting the particular geometries for which the onset of the racemic-scalemic transition will precede the Alder-Wainwright transition. Because the roots of equation (51) are determined primarily by the singularity of the Pavlícek-Nezbeda-Boublík equation of state, (25), one might reasonably expect that the racemic-scalemic transition will most easily occur for particles of the greatest degrees of asphericity and chirality, for a given molecular volume. In any case, whenever the racemic fluid mixture has a positive excess volume, one should reasonably expect the same property for the solid phase and that, when the thermodynamic chirality function is greater than the threshold value, as given by equations (13) and (21), the fluid will solidify as mixed crystals, *not* as a solid solution, - as Pasteur observed in 1848.

In the formal analysis of hard-body fluids in section 2, the chiral properties of the components was manifested through their excess volume for reason of the geometric differences described by equations (24). In general, the property of chiral hard bodies can be set forth as:

$$\left.\begin{array}{l} \tilde{R}_D = \tilde{R}_L \quad \tilde{S}_D = \tilde{S}_L \quad \tilde{V}_D = \tilde{V}_L \\ \text{and} \\ \tilde{R}_D \tilde{S}_D = \tilde{R}_L \tilde{S}_L \\ \text{but} \\ \tilde{R}_D \tilde{S}_D \neq \tilde{R}_L \tilde{S}_D = \tilde{R}_D \tilde{S}_L \end{array}\right\}. \qquad (53)$$

The properties set forth by equations (53) result automatically by virtue of the strict satisfaction of equations (11) which assures that the excess volume is attributable solely to the chirality. However, in a case of "closely-similar, but achiral, molecules, for which $\boldsymbol{m}_1^\varnothing \approx \boldsymbol{m}_2^\varnothing$, or $V_{m,1} \approx V_{m,2}$, or both, there could exist the circumstance for which

$$\left.\begin{array}{l} \tilde{R}_1 \approx \tilde{R}_2 \quad \tilde{S}_1 \approx \tilde{S}_2 \quad \tilde{V}_1 \approx \tilde{V}_2 \\ \text{and} \\ \tilde{R}_1 \tilde{S}_1 \approx \tilde{R}_2 \tilde{S}_2 \\ \text{but also} \\ \tilde{R}_1 \tilde{S}_1 \neq \tilde{R}_2 \tilde{S}_1 \end{array}\right\}. \quad (54)$$

A moment's consideration of the Pavlícek-Nezbeda-Boublík equations assures that such a system could also undergo phase separation, or (if possible) molecular conversion, analogous to the racemic-scalemic transition at high density. For such, the pressure at which the thermodynamic Affinity changes sign, as in Fig. 5, would simply be shifted. Thus the present formal analysis of hard-body systems subsumes the cases investigated by Vlot *et al.*[40-42] by molecular simulation.

This property just described perhaps states a very important aspect of the phenomenon of optical activity in fluids: **The racemic-scalemic transition is only *one particular* example of the general, complex phase behavior characteristic of "closely-similar" molecules at high pressure.** The racemic-scalemic transition, or the onset of optical activity in fluids, is a straight-forward phase-stability problem in chemical thermodynamics.

In summary: This analysis has demonstrated that the evolution of optically-active multicomponent systems is an inevitable consequence of the universal geometric properties of the directional, covalent bond, manifested in accordance with the fundamental dictates of thermodynamic stability theory. This analysis of the genesis of such of optically-active systems has required neither invocation of any *deus ex machina* such as "panspermia," nor any (highly questionable) suggestions of some amplification of the parity non-conserving weak interaction involved in beta decay. The chiral property of certain *molecules* is simply the inevitable consequence of the directional nature of the covalent chemical bond, - as characterized by, for example, the single-branched alkanes. The optical activity of abiotic *fluid systems* is an inevitable consequence of the effect of the thermodynamic excess volume upon the Gibbs free enthalpy, particularly at high density, - as characterized by, for example, natural petroleum. The racemic-scalemic transition is simply one special example the complex phase behavior manifested at high densities by multicomponent systems containing "nearly-similar" molecules.

Note added in proof: The authors have just become aware of two prescient early papers by Scott which take up the thermodynamic problem of the stability of mixtures of enantiomers.[43, 44] Although these papers do not analyze the details of the molecular properties which determine whether phase separation will occur (i.e., the magnitude of the thermodynamic chirality functional), nor discuss enantiomeric conversion processes, nor predict the racemic-scalemic transition, they do predict the phase separation of enantiomers. In his 1987 paper, Scott remarked, "… one has to find fluid-fluid phase separation in racemic mixtures." With the Monte Carlo investigations of fluorochloroiodomethane and 4-vinylcyclohexene described in section 4, such has now been done.

**Appendix I.    Concerning the optical activity observed in systems of biotic molecules.**

Although this article has addressed the phenomenon of the evolution of optical activity in abiotic fluid systems, because of the general interest in fluids containing molecules of biotic origin, at least a few words are in order concerning the significance of the present results for such. As has been pointed out, the racemic-scalemic transition depends upon the molecular volume and degree of asphericity; and the greater of each, the more susceptible is the system to the racemic-scalemic transition. The biotic molecules, including particularly such as DNA or cellulose, are very large molecules, and aspherical; both DNA and cellulose are also chiral. The estimated threshold value of the binary thermodynamic chirality function, $Q \approx 2$, requires that for $T \approx 300$ and $p = 1$ bar, a molar excess volume of approximately $V^E \approx 4.8 \cdot 10^3$ cm$^3$/mol. The size of, for example, a typical protein of approximately 200 units, is approximately $1\text{-}5 \cdot 10^{-10}$ cm$^3$/molecule;[45] and its molar volume is thus greater than approximately $1.2 \cdot 10^6$ cm$^3$/mol. The excess volume of typical mixtures of unlike molecules is on the order of a few tenths of a percent to a few percent of their individual molar volumes. Thus, even a very small degree of chirality will endow a protein, or any similarly large biotic molecule, with a molar excess volume for which the entropy component attributable to the thermodynamic chirality function will overwhelm the entropy of mixing, at *low* pressures and *all* temperatures for which such molecules exist. Therefore, the phenomenon of optical activity in systems of biotic molecules deserves to be considered simply as a thermodynamic inevitability.

(Note: Consideration of why terrestrial biotic systems manifest entirely pure-component abundances of only a single enantiomer, and always that of a single chirality, - as, e.g., in L-DNA, - is beyond the scope of this present article. Such phenomena will be addressed in a future article in which will be taken up the subjects of phase stability, population dynamics, and their interaction, of large chiral molecules.)

**Appendix II.** **Concerning the optical activity observed in material extracted from meteorites.**

The Introduction to this article refers to observations of optical activity in material extracted from carbonaceous meteorites, and to the decisive role which such played in the recognition that optical activity has no intrinsic connection with biotic matter or processes. For that reason, a few words are appropriate concerning the significance of the results of this present analysis for the previous ones of meteorite material.

Analysis of the material in carbonaceous meteorites has established that their minerals, rocks, and agglomerated material went through various stages of fractionation in the course of formation. From their chemical composition and calculated velocity of cooling, the carbonaceous chondrites have been ascertained to be high-pressure remnants, originally formed at depths between 70-150 km in parent celestial bodies whose diameters were more than 800 km, and perhaps as great as 2500 km.[46] The observations not only of heavy hydrocarbon molecules but also of diamond minerals, in their cubic (diamond), hexagonal (lonsdaleite), and mixed (chaoite) forms, in such as the Novo Urei, Goalpara, and North Haig meteorites, attest to their high-pressure history.[47]

Thus the present thermodynamic analysis of optical activity, which has invoked no specific material properties except the necessary one of molecular chirality, is consistent with the astrophysical and chemical analysis of materials extracted from meteorites. The results here reported both confirm and support the previous analyses of the material from meteorites.

# References:

1    L. Pasteur, "Sur la dissymétrie moleculaire," *C.R. Hebd. Séanc*, 1848, **26**, 535.
2    M. H. Engel and B. Nagy, "Distribution and enantiomeric composition of amino acids in the Murchison meteorite," *Nature*, 1982, **296**, 837-840.
3    M. H. Engel, S. A. Macko and J. A. Silfer, "Carbon isotope composition of individual amino acids in the Murchison meteorite," *Nature*, 1990, **348**, 47-49.
4    M. H. Engel and S. A. Macko, "Isotopic evidence for extraterrestrial non-racemic amino acids in the Murchison meteorite," *Nature*, 1997, **389**, 265-268.
5    B. Nagy, "Optical Activity in the Orgueil meteorite," *Science*, 1965, **150**, 1846.
6    B. Nagy, *Carbonaceous Meteorites*, Elsevier, Amsterdam, 1975.
7    S. Arrhenius, *Worlds in the Making: The Evolution of the Universe*, Harper, New York, 1908.
8    F. H. Crick and L. E. Orgel *Icarus*, 1973, **19**, 341.
9    B. Mason, "Meteorites," *American Scientist*, 1957, **55**, 429-455.

10  F. Vester, T. Ulbricht and H. Krauch *Naturwissenschaften*, 1959, **46**, 68.
11  T. Ulbricht *Quart. Rev.*, 1959, **13**, 48.
12  T. Ulbricht *Origins of Life*, 1975, **6**, 303.
13  T. Ulbricht in *Comparative Biochemistry*, ed. M. Florkin and H. S. Mason, Academic Press, New York, 1962, vol. IV, Part B, 16-25.
14  S. F. Mason, "Origin of biomolecular chirality," *Nature*, 1985, **314**, 400-401.
15  S. F. Mason and G. E. Tranter, "*Electroweak enantiomeric inequivalences: Origins of biomolecular handedness*," ed. B. Kurtev, 1987, 196-217.
16  S. F. Mason and G. E. Tranter, "The parity-violating energy difference between enantiomeric molecules," *Chem. Phys. Lett.*, 1983, **94**, 34-37.
17  J. Jacques, A. Collet and S. H. Wilen, *Enantiomers, Racemates, and Resolutions*, John Wiley & Sons, New York, 1981.
18  E. A. Guggenheim, "The theoretical basis for Raoult's law," *Trans. Faraday Soc.*, 1937, **33**, 151-159.
19  G. Scatchard, "Equilibrium in non-electrolyte mixtures," *Chem. Rev.*, 1949, **44**, 7-35.
20  R. M. Corless, G. H. Gonnet, D. E. G. Hare, D. J. Jeffrey and E. Donald, "On the Lambert W function," *Adv. Comp. Math.*, 1996, **5**, 329-359.
21  R. M. Corless, D. J. Jeffrey and D. E. Knuth, "A sequence of series for the Lambert W function," in *Proc. A.S.S.A.C.*, ed. W. Kuchlin, 1997, Maui.
22  J. S. Rowlinson and F. L. Swinton, *Liquids and Liquid Mixtures*, Butterworth Scientific, London, 1982.
23  U. K. Deiters, M. Neichel and E. U. Franck, "Prediction of the thermodynamic properties of hydrogen-oxygen mixtures from 80 to 373 K and to 100 MPa," *Ber. Bunsenges. Phys. Chem.*, 1993, **97**, 649-657.
24  H. Reiss, H. L. Frisch and J. L. Lebowitz, "Statistical mechanics of rigid spheres," *J. Chem. Phys.*, 1959, **31**, 369-380.
25  H. Reiss, H. L. Frisch and J. L. Lebowitz, "Mixtures of hard spheres," in *The Equilibrium Theory of Classical Fluids*, ed. H. L. Frisch and J. L. Lebowitz, W. A. Benjamin, New York, 1964, II-299 - II-302.
26  J. Pavlícek, I. Nezbeda and T. Boublík, "Equation of state for hard convex fluids," *Czech. J. Phys.*, 1979, **B29**, 1061-1070.
27  T. De Donder, *L'Affinité*, Gautier-Villars, Paris, 1936.
28  D. Kondepudi and I. Prigogine, *Modern Thermodynamics: From Heat Engines to Dissipative Structures*, John Wiley & Sons, New York, 1998.
29  I. Prigogine and R. Defay, *Chemical Thermodynamics*, Longmans, London, 1954.
30  I. Prigogine, *Nonequilbrium Statistical Mechanics*, John Wiley - Interscience, New York, 1962.
31  J. S. Rowlinson, *Liquids and Liquid Mixtures*, Butterworth, London, 1959.
32  J. F. Kenney, "The evolution of multicomponent systems at high pressures: III. The effects of particle shape upon the Alder-Wainwright transition in hard-body gases," *Phys. Chem. Chem. Phys.*, 1999, **1**, 4323-4327.
33  N. L. Allinger, Y. H. Yuh and J.-H. Lii, "Molecular mechanics: The MM3 force field for hydrocarbons. 1.," *J. Amer. Chem. Soc.*, 1989, **111**, 8551-8566.

34  J.-H. Lii and N. L. Allinger, "Molecular mechanics: The MM3 force field for hydrocarbons. 2. Vibrational frequencies and thermodynamics," *J. Amer. Chem. Soc.*, 1989, **111**, 8566-8575.
35  J.-H. Lii and N. L. Allinger, "Molecular mechanics: The MM3 force field for hydrocarbons. 3. The van der Waals potentials and crystal data for aliphatic and aromatic compounds," *J. Amer. Chem. Soc.*, 1989, **111**, 8575-8582.
36  J. Emsley, *Die Elemente*, De Gruyter, Berlin, 1994.
37  L. P. N. Rebelo, "A simple $g^E$-model for generating all basic types of liquid-liquid equilibria and their pressure dependence. Thermodynamic constraints at critical loci," *Phys. Chem. Chem. Phys.*, 1999, **1**, 4277-4286.
38  "Symposium on Occurrence of Petroleum in Igneous and Metamorphic Rocks," *American Association of Petroleum Geologists Bulletin*, 1932, **16**, 717-858.
39  J. F. Kenney, "The evolution of multicomponent systems at high pressures: II. The Alder-Wainwright, high-density, gas-solid phase transition of the hard-sphere fluid," *Phys. Chem. Chem. Phys.*, 1999, **1**, 3277-3285.
40  M. J. Vlot, S. Claassen, H. E. A. Huitema and J. P. van der Eerden, "Monte carlo simulations of racemic liquid mixtures: Thermodynamic properties and local structure," *Mol. Phys.*, 1997, **91-30**, 19.
41  M. J. Vlot, H. E. A. Huitema, A. de Vooys and J. P. van der Eerden, "Crystal structures of symmetric Lennard-Jones mixtures," *J. Chem. Phys.*, 1997, **107**, 4345-4349.
42  M. J. Vlot, J. C. van Miltenburg and H. A. J. Oonk, "Phase diagrams of scalemic mixtures: A monte carlo simulation study," *J. Chem. Phys.*, 1997, **107**, 10102-10111.
43  R. L. Scott, "Models for phase equilibria in fluid mixtures," *Accounts of Chemical Research*, 1987, **20**, 97-107.
44  R. L. Scott, "Modification of the phase rule for optical isomers and other symmetric systems," *J. Chem. Soc. Faraday II*, 1977, **73**, 356-360.
45  A. Y. Grosberg and A. R. Khokhlov, *Giant Molecules - Here, There, Everywhere*, Academic Press, London, 1997.
46  G. Mueller, "Interpretation of the microstructure of carbonaceous meteorites," in *Advances in Organic Geochemistry*, Pergamon Press, London, 1964, 119-140.
47  G. P. Vdovykin, "Ureilites," *Space Science Rev.*, 1970, **10**, 483-510.